\documentclass[12pt]{article}
\usepackage{epsfig,amsmath,amsfonts,amssymb,setspace,multirow,textcomp, siunitx}
\usepackage[T1]{fontenc}
\usepackage[dvipsnames]{xcolor}
\begin{document}
\title{X-ray galaxies selected from HyperLEDA database}
\author{\textsl{ N.~Pulatova$^{1,3}$, A.~Tugay$^{2}$, L.~Zadorozhna$^{2,4}$,} \\ R. Seeburger $^{1}$,  	  O.~Gugnin$^{2}$ }
\maketitle
\begin{center} {\small 
$^{1}$
Max-Planck-Institut f\"ur Astronomie, K\"onigstuhl 17, D-69117 Heidelberg, Germany\\
$^{2}$Taras Shevchenko National University of Kyiv, Hlushkova ave, 4, 03127, Kyiv, Ukraine\\
$^{3}$Main Astronomical Observatory of the National Academy of Sciences of Ukraine, Akademika Zabolotnoho str, 27, 03143, Kyiv, Ukraine\\

$^{4}$Jagiellonian University, Faculty of Physics, Astronomy and Applied Computer Science, ul. prof. Stanis{{\l}}awa {{\L}}ojasiewicza 11, 30-348, Krak{{\'o}}w  \\

{\tt nadiia.pul@gmail.com}}

\end{center}
\begin{abstract}
 We cross-matched the 4XMM-DR10 catalog with the HyperLEDA database and obtained the new sample of galaxies that contain X-ray sources. Excluding duplicate observations and false matches, we present a total of 7759 galaxies with X-ray sources. In the current work, we present general properties of the sample: namely the distribution in equatorial coordinates, radial velocity distribution, morphological type, and X-ray fluxes. The sample includes morphological classification for 5241 galaxies with X-ray emission, almost half of which, 42\%, are elliptical (E, E-S0). Most galaxies in the sample have nuclear X-ray emission (6313 or 81\%), and the remaining 1443 (19\%) present X-ray emission from the host galaxy. This sample can be used for future deep studies of multi wavelengths properties of the galaxies with X-ray emission.
\\[1ex]
{\bf Key words:} X-rays: galaxies, catalogues, galaxies: statistics
\end{abstract}
\section*{\sc introduction}
The X-ray emission of galaxies is theorized to arise mostly from accretion on the supermassive black holes for active galactic nuclei (AGN) \cite{lynden69, 2012AlexanderHickox, 2005BrandtHasinger, 2020Vasylenko, 2015Pulatova, 2016Vavilova}, onto compact objects in binary systems, and from the hot phase of the interstellar medium for non-AGN \cite{basu20, 2012Fabbiano}. Therefore, X-ray observations of galaxies's central regions allow us to study properties of accretion disc around compact objects. 
 The broad studies of the numerous galaxies allow us to detect general patterns in the distribution of their parameters or investigate a certain type of their nuclear activity \cite{gali21}.

A recent study \cite{zador21} compiled a catalog of 1172 X-ray galaxies. Each of the galaxies from this catalog was manually verified. There are a series of works, dedicated to XXL region, that was observed with the XMM Newton telescope \cite{2012Pierre, 2016Menzel, 2018Melnyk}. The authors detected in XMM XXL field 50 000 X-ray AGNs, calculated correlation function for each of them, and investigated their environment. Multiple images from XXL field among the optical counterparts of X-ray selected point-like sources were analyzed in \cite{2015Finet}, and the number and statistics was calculated. In \cite{2012Elyiv} the authors have found that the sources with a hard-spectrum are more clustered than soft-spectrum sources, which may indicate that the two main types of AGN populate different environments. The influence of environment was studied more deeply in \cite{2018Koulouridis, 2015Vavilova, 2015aVavilova}. 
The results of this investigation provide evidence of the physical mechanisms that drive AGN and galaxy evolution within clusters.

 According to detailed searching for every galaxy from the catalog, it was found found that most galaxies have an active X-ray nucleus; Seyfert galaxies dominate at short distances, and quasars are prevalent at large distances. It was revealed 169 galaxies exhibit extended nuclei with a visible surface brightness distribution and 173 galaxies with more than one X-ray source. From the 1172 X-ray galaxies of the catalog we selected 11 nearby optically bright QSOs \cite{2021Zadorozhna}. More than half of the sources (6 out of 11) are classified as radio-quiet QSOs. These sources have steep spectra $\Gamma > 2.1$. Extremely steep photon index ($\sim 2.4 - 2.5$) occurs for three radio-loud type I quasars.

Together with Chandra NuSTAR, Fermi-LAT, eROSITA X-ray observatories, XMM-Newton is one of the most important sources of X-ray observations. It is a space observatory that has been observing the sky in X-ray band for the last 20 years. The 4XMM-DR10 catalog contains data from 11647 observations covering an energy interval from 0.2 keV to 12 keV, which corresponds to 575158 unique X-ray sources \cite{webb20}. HyperLEDA\footnote{http://leda.univ-lyon1.fr/} is a database to study the physics of the galaxies and cosmology, which combines optical measurements published in literature and modern surveys into a unique homogeneous description of the astronomical objects \cite{makar14}. 

Samples of X-ray galaxy clusters are used in observational cosmology for the derivation of cosmological parameters. X-ray observations of galaxy clusters are also used for the studies of galaxy clusters themselves, the physics of intracluster gas, interaction processes, matter transportation, and feedback processes. \cite{kirk21}. In particular, properties of X-ray galaxy clusters were studied in the following papers: \cite{2020TugayShevch, 2016TugayDylda, 2017ShevchTugay, 2018Tugay}.
\section*{\sc sample of x-ray galaxies}

In this work we chose the 4XMM-DR10 as one of the largest repository of X-ray sources from a single space telescope \cite{webb20}. We cross-matched XMM-Newton catalog with the HyperLEDA database. The HyperLEDA database contains more than 1.5 million objects that are identified as galaxies covering whole sky.  The 4XMM-DR10 database covers 849 991 detections of 575 158 unique sources over 2.85\% of the sky. The fluxes and spectra are accessible for 303023 bright sources (53\% of all observations). We performed automatic cross-matching of these two databases using TOPCAT, and created the sample which includes 7905 X-ray galaxies. 
Correlation was performed in two stages. First, since the EPIC cameras of the XMM the telescope's has field of view (FOV) of $30'$ and in the energy range from 0.15 to 15 keV and angular resolution ($6.6''$ the FWHM of the
point spread function (PSF) \cite{Struder2001}), a galaxy in the HyperLEDA base may be identified as an X-ray source when the latter is found
among the 4XMM-DR10 objects at a distance smaller than $6.6''$ from the galaxy. We rounded X-ray resolution and choose the limit $7''$ between X-ray source and optical center of a galaxy. The second stage was to determine the upper
boundary for the number of X-ray galaxies, we increased the radius of correlation between the X-ray 
4XMM-DR10 source and a galaxy`s center from the HyperLEDA database to a value of the semimajor axis of the galaxy
in the u spectral band (the SDSS system). The u band was used because the number of galaxies, the sizes
of which are available, is largest in this band. This extension of the radius resulted in detecting extragalactic X-ray sources beyond the galactic nuclei. Some of the input data was false or repetitive. Therefore, we performed checking for duplicate observations and having false matches for spatially close objects. The several closest galaxies with large angular sizes were excluded from the sample because of high probability of containing the background X-ray sources that are not connected with the host galaxy. After this we determined  272 objects in our sample which were excluded as double entities. We excluded 3 galaxies with angular size more $26'$ (ESO056-115, ESO206-020A, NGC5457) containing many background X-ray sources not associated with the goal of our research. Our sample was reduced  from 7905 objects to 7759 galaxies emitting X-rays. 
For a few close galaxies, we have multiple (two or more) X-ray sources.  We studied them in our previous paper, published in 2021 \cite{zador21}. In this paper we concentrate on distant galaxies with X-ray emission. If there was more than one X-ray source in a galaxy, we chose the brighter source located closer to the galactic center.

For each galaxy we include the following parameters: Right ascension ($deg$), Declination ($deg$), Optical B magnitude ($mag$), Optical size ($arcsec$) from the HyperLEDA database, Radial velocity cz ($km/s$), X-ray flux in total photon energy band (0.2 - 12 keV) from the XMM-DR10 catalog($erg/cm^2 \cdot s$), the angular separation between the position of galaxy's center in the HyperLEDA and XMM-Newton catalogues ($arcsec$), and the designation of each object according to the Catalogue of Principal Galaxies (PGC). Morphological types are given for 67\% of the sample. The rest of the sample consists of distant galaxies which appear as point sources. 

 Our sample of X-ray galaxies, which were identified in the cross-matching of these two databases, may form a basis for deeper studies for sampling certain interesting groups of objects, building spectra of objects of certain classes, and developing or improving the theory of their emission. 

\section*{\sc results}

We present the general properties of  distributions of the sample`s parameters: the distribution of galaxies in equatorial coordinates, distributions of morphological types, the angular separation between the position of galaxy's center in the HyperLEDA and XMM-Newton catalogs, and angular diameter of a galaxy. Additionally, we discuss the relation between optical magnitude and X-ray flux, and the relation between optical magnitude and radial velocity for the galaxies with X-ray sources in the sample. 

\bigskip
\textbf{1. Distribution of the sample of galaxies with X-ray sources in equatorial coordinates}
\smallskip

\begin{figure}[htp]
\centering
\includegraphics[width=0.9\textwidth]{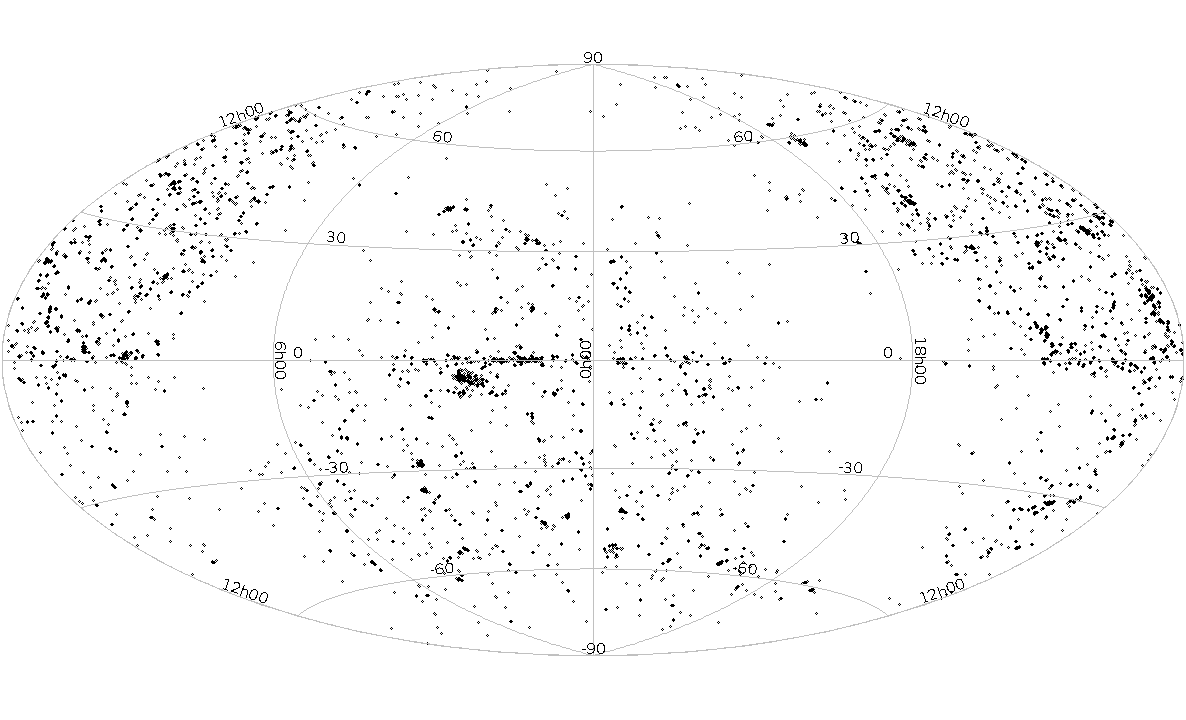}
\caption{Distribution in equatorial coordinates}
\label{fig1}
\end{figure}

Figure \ref{fig1} shows the distribution of all galaxies of the sample on the celestial sphere with equatorial coordinate system. It is possible to infer from the figure that X-ray galaxies of the sample tend towards a homogeneous distribution except the zones in the vicinity of 6h and 18h right ascensions. They correspond to the zone of avoidance of the HyperLEDA \cite{makar14}. At the same time, the observations of XMM-Newton have close to uniform distribution on the celestial sphere \cite{webb20}. Some small regions with higher density of detected galaxies with X-ray emission do not indicate higher spacial density of galaxies with X-ray sources. This feature corresponds to the most intensive observations with XMM-Newton in these regions. 

\bigskip

\textbf{2. Morphological classification of the sample of galaxies with X-ray sources} 
\smallskip

Out of 7759 galaxies with X-ray sources, there are 5241 galaxies for which the morphological type is given in HyperLEDA database (67\% of the full sample). Almost half of the galaxies with X-ray emission with morphological classification are elliptical (E, E-S0) 42\%, 35\% as spiral without bars (S0-a, S0, Sb, Sab, Sa, Sm), 12\% as barred galaxies (SBc, SBb, SBab, SBbc, SBa, SBm, SBcd, SBd), 2\% as spiral galaxies with weak bars (SABb, SABc, SABa, SABd), and the other 1\% as irregular (I, IB, IAB). Table \ref{table1} shows the exact number of galaxies of each type in the sample. A question mark next to the type designation indicates that classification of these galaxies is not clear.  

\begin{table}[h]
 \centering
 \caption{Morphological classification}\label{table1}
 \vspace*{1ex}
  \begin{tabular}{|c|c|c|c|c|c|c|c|c|c|}
  \hline

\hline
Type & E    & E-S0   	& E?   	& S0-a 	& S0   	& S?   	& Sb 	& Sab 	& Sa   \\	   	
Number & 1706 & 279	& 195	& 525	& 409	& 318	& 254	& 174	& 148 \\			
\hline
Type & Sm & Sc	& SBc & 	SBb & Scd-Sm & SBab 	& SBbc 	& SBa 	& SBm \\
Number & 18 & 340	& 306 & 101 & 75	& 68	& 66	& 55	& 12\\
\hline
Type 	& SBcd 	& SBd  	& SABb 	& SABc 	& SABa	& SABd 	& I    	& IB   	& IAB 	\\
Number	& 13	& 7	& 42	& 34	& 29	& 6	& 42	& 14	& 5	\\

\hline
\end{tabular}
\end{table}

\bigskip

\textbf{3. The angular separation between the position of objects in HyperLEDA and XMM-Newton catalogues}
\smallskip

\begin{figure}[htp]
\centering
\includegraphics[width=0.8\textwidth]{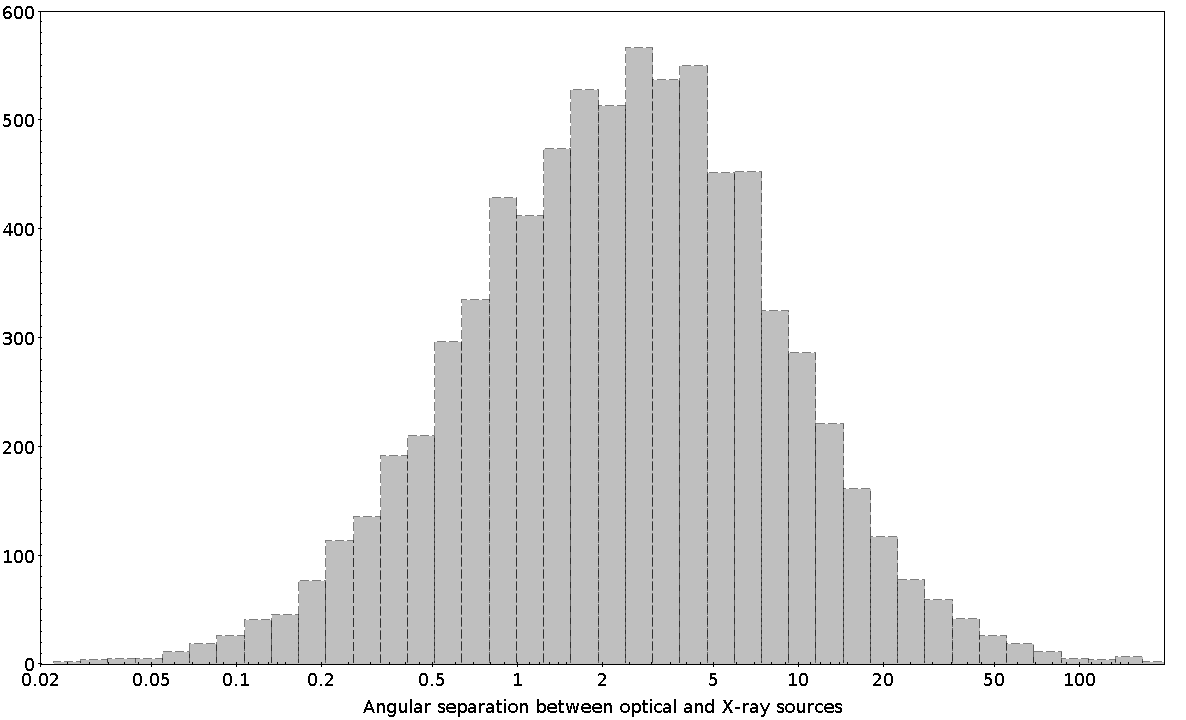}
\caption{The angular separation between the position of objects 
in HyperLEDA and XMM-Newton catalogues, in $arcsec ('')$.}
\label{fig2}
\end{figure}

Figure \ref{fig2} shows the distribution of the angular separation between the position of objects in HyperLEDA and XMM-Newton catalogs in arcseconds. The minimum value of distance is \ang{;;0.01}, the maximum value is \ang{;;191}, and the mean value is \ang{;;6.7}. It is worth noticing that the mean value corresponds to the angular resolution of XMM-Newton observations. 6314 galaxies of our sample have separation between X-ray source and the optical center of a galaxy smaller than \ang{;;7}. We put a limit for angular separation between the position of objects in HyperLEDA and XMM catalogs that corresponds to the resolution of XMM telescope. It was chosen with the idea that even if 2 or more X-ray sources exist in this area, we will be able to identify only one common X-ray source. But we also included to whole sample X-ray sources which belongs to the host galaxy. 
The shape of this distribution is similar to the logarithmic Gaussian. In this case, the falling inflection point is about 6''. After this point the number of detected objects decreasing rapidly. Most of galaxies of the sample belong to galaxies with X-ray nuclear emission, namely 6313 galaxies or 81\% of whole sample, and 1443 galaxies (19\%) present X-ray emission from the host galaxy.

\bigskip

\textbf{4. Angular size distribution}
\smallskip

\begin{figure}[htp]
\centering
\includegraphics[width=0.9\textwidth]{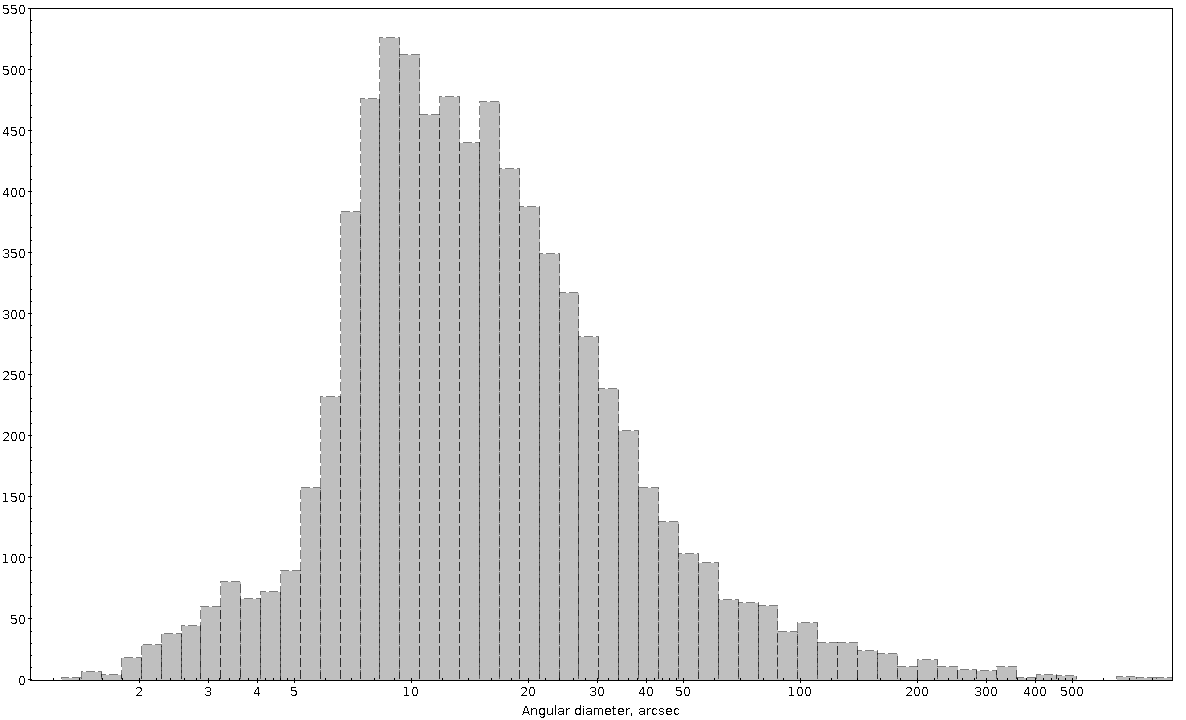}
\caption{Distribution by angular diameter, in $arcsec ('')$}
\label{fig3}
\end{figure}

Figure \ref{fig3} shows the distribution of optical large semiaxis $a$ of galaxies in arcseconds. The minimum value of $a$ is \ang{;;1.3}, the maximum value is $25.6'$ corresponding to NGC5128.  The mean value of angular size is \ang{;;25.8}.

\bigskip


\textbf{5. Optical B magnitude and X-ray flux}
\smallskip

\begin{figure}[htp]
\centering
\includegraphics[width=0.8\textwidth]{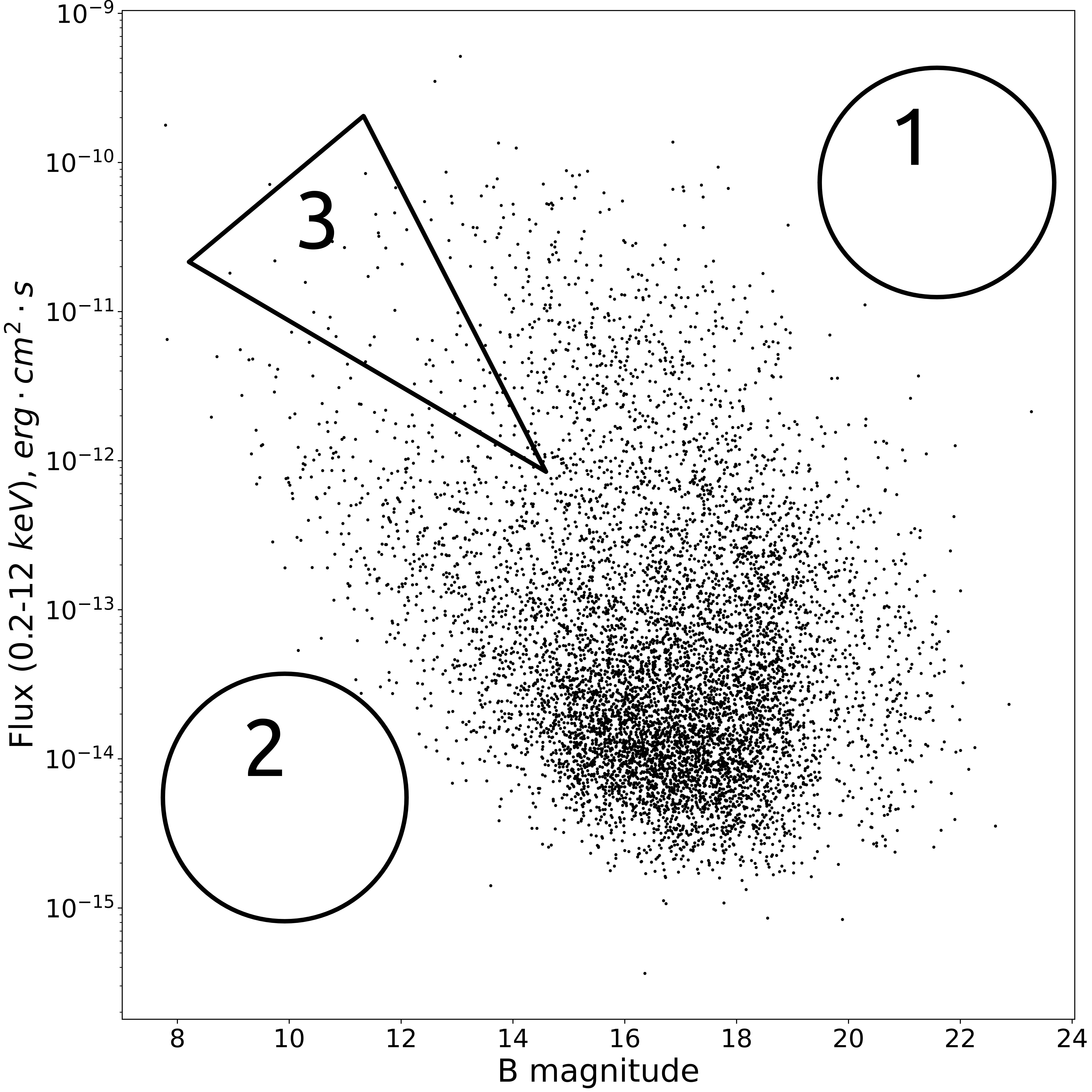}
\caption{Optical B magnitude and X-ray flux. The three regions marked with the numbers 1-3 corresponds to: 1 - brightest galaxies with low redshift; 2 - the limit of  XMM sensitivity; 3- the gap between two populations of active galaxies.}
\label{fig4}
\end{figure}

Figure \ref{fig4} shows the relation between X-ray flux in total photon energy band (0.2 - 12 keV) in $erg/cm^2\cdot s$ and the optical  magnitude for the sample of galaxies with X-ray sources. The optical magnitude is given in B band. The bottom part of the diagram corresponds to the limit of the instrument's sensitivity (region number 2 on Figure \ref{fig4}). 
On this diagram there are some galaxies with X-ray sources bright in both X-ray and optical; they are low redshift galaxies (region number 1 on Figure \ref{fig4}). Most of the galaxies with faint X-ray sources are also faint in the optical.
Another feature that we see that this diagram demonstrates two  populations: bright in X-ray sources and faint optical galaxies and bright optical and faint X-ray sources. The upper right group is more active galaxies, the bottom left is less active objects and the groups are separated by a gap (region number 3 on Figure \ref{fig4}). Most of the galaxies are in the range of B magnitude 15--20 $mag$ and X-ray fluxes $10^{-14}-10^{-13} erg/cm^2 \cdot s$.

\bigskip

\textbf{6. Optical B magnitude and radial velocity cz}
\smallskip

\begin{figure}[htp]
\centering
\includegraphics[width=0.8 \textwidth]{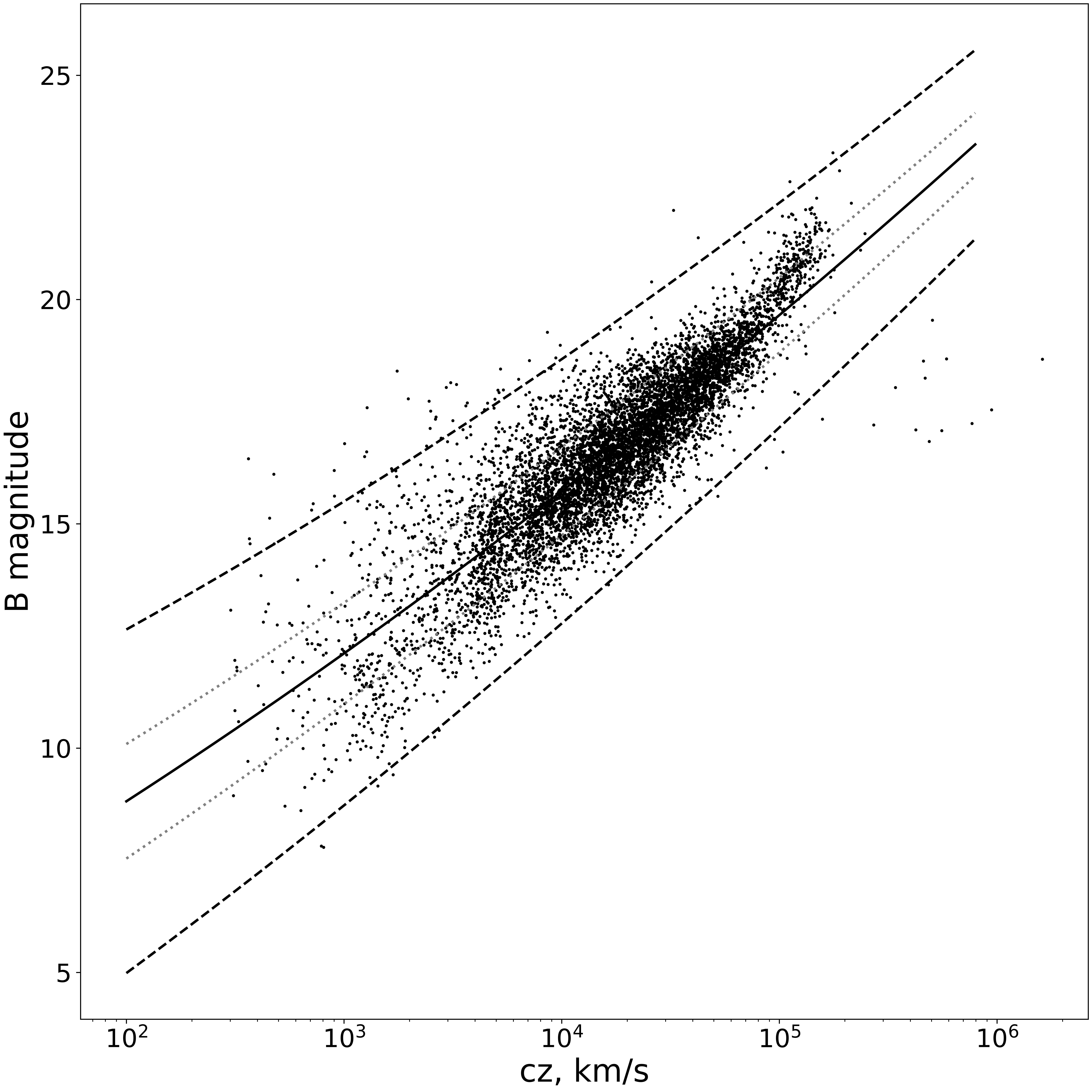}
\caption{Optical B magnitude and radial velocity cz. The solid line represents the best fit $cz=a\cdot B^2 + b \cdot B + c$, where: $cz$, $B$ - radial velocity and B magnitude; $a, b, c$ - free parameters. The values $a=0.16$; $b=2.51$; $c=3.17$ was obtained by applying "optimize" function from "scipy" package. Gray dotted and black dashed lines represents 1 $\sigma$ and 3 $\sigma$ error respectively }
\label{fig5}              
\end{figure}

Figure \ref{fig5} shows the relation between radial velocity in km/s and optical magnitude. Optical magnitude is given in B band. Radial velocity was calculated from the redshift using $v=cz$ dependence. Most galaxies follow a linear trend and lay in the range of 15-20  $mag$ and $10^{4}-10^{5}$ $km/s$. 
The brightest galaxies are located in the upper part of the graph due to having high value of velocity (>$2\cdot 10^5$ kms) and low magnitude (around $18^m$). We suppose that these observational features are occurrences of the selection effect, because the sensitivity of telescopes does not allow detection of fainter objects at higher distances (and higher $cz$ respectively). These galaxies are beamed quasars, namely the jets of beamed quasars that are oriented to the observer; therefore there is high intensity optical emission \cite{Wang2006A, Schwartz2006}. 

\section*{\sc conclusions }
                          
We created the sample of  7759 X-ray galaxies from XMM Newton DR10 catalog identified with HyperLEDA database. Equatorial coordinates, optical magnitude, angular size, radial velocity, X-ray flux, distance between optical and X-ray images, and name are given for each X-ray galaxy. Most galaxies in the sample have nuclear X-ray emission (6313 or 81\%), and the remaining 1443 (19\%) present X-ray emission from the host galaxy. We found and analyzed properties of parameter distributions of the sample, which can be used for future studies of X-ray properties of
the galaxies. The whole sample is available at our website\footnote{https://sites.google.com/view/xgal}.

\section*{\sc acknowledgements}
The authors are very grateful to Dr Morgan Fouesneau (Max-Planck-Institut f\"ur Astronomie) for  modelling "radial velocity versus B magnitude" with a quadratic regression and to  Mr
Alexander J. Dimoff (Max-Planck-Institut f\"ur Astronomie) for the helpful suggestion which has improved the English text of our paper. Nadiia G. Pulatova is grateful for the financial support of this work to The Max Planck Society, Germany and to director of the Max-Planck-Institut f\"ur Astronomie Prof. Dr. Dr. h.c. Thomas K. Henning for supporting Ukrainian astronomers. The work of Lidiia V. Zadorozhna was supported by the grant No. UMO-2018/30/Q/ST9/00795 from the National Science Centre, Poland. 
This research had been made with the support of the Center for the Collective Use of
Scientific Equipment "Laboratory of high energy physics and astrophysics".
This research has made use of data obtained from the 4XMM XMM-Newton serendipitous source catalogue compiled by the 10 institutes of the XMM-Newton Survey Science Centre selected by ESA.

We acknowledge the usage of the HyperLEDA database. 

Funding for the Sloan Digital Sky 
Survey IV has been provided by the 
Alfred P. Sloan Foundation, the U.S. 
Department of Energy Office of 
Science, and the Participating 
Institutions. SDSS-IV acknowledges support and 
resources from the Center for High 
Performance Computing  at the 
University of Utah. The SDSS 
website is www.sdss.org.


\begin{thebibliography} {7}
{\small
\bibitem{webb20} Webb~N.\,A., Coriat~M., Traulsen I. et al. 2020, A\&A, 641, A136. https://doi.org/10.1051/0004-6361/201937353

\bibitem{makar14} Makarov~D., Prugniel~P., Terekhova~N., Courtois H., Vauglin I. 2014, A\&A, 570, A13. https://doi.org/10.1051/0004-6361/201423496

\bibitem{basu20} Basu-Zych~A.\,R., Hornschemeier~A.\,E., Haberl~F. et al. 2020, MNRAS, 498, 1651. https://doi.org/10.1093/mnras/staa2481

\bibitem{gali21} Galiullin~I., Gilfanov~M. 2021, A\&A 646, A85. https://doi.org/10.1051/0004-6361/202039522

\bibitem{zador21} Zadorozhna~L.\,V., Tugay~A.\,V., Shevchenko~S.\,Y., Pulatova~N.\,G. 2021, Kinematics and Physics of Celestial Bodies, Vol. 37, No. 3, pp. 149-157. https://doi.org/10.15407/kfnt2021.03.068

\bibitem{lynden69} Lynden-Bell~D. 1969, Nature, vol. 223, 690-694. https://doi.org/10.1038/223690a0

\bibitem{kirk21} Kirkpatrick~C.\,C., Clerc~N., Finoguenov~A. 2021, MNRAS, 503, 5763-5777 https://doi.org/10.1093/mnras/stab127
}
\bibitem{2012AlexanderHickox} Alexander, D.~M. \& Hickox, R.~C.\ 2012, New Astronomy Reviews, 56, 93. https://doi.org/10.1016/j.newar.2011.11.003

\bibitem{2005BrandtHasinger} Brandt, W.~N. \& Hasinger, G.\ 2005, Annual Review of Astronomy \& Astrophysics, 43, 827. https://doi.org/10.1146/annurev.astro.43.051804.102213

\bibitem{2012Fabbiano} Fabbiano, G.\ 2012, Astrophysics and Space Science Library, 378, 1. https://doi.org/10.1007/978-1-4614-0580-1\_1

\bibitem{2021Zadorozhna} Zadorozhna L. V.,  Tugay A. V.,  Maluy O. I.,  Pulatova N. G.\ 2021, Journal of physical studies, 25, 4901. https://doi.org/10.30970/jps.25.4901

\bibitem{2020TugayShevch} Tugay, A.~V., Zadorozhna, L.~V., \& Shevchenko, S.~Y.\ 2020, Odessa Astronomical Publications, 33, 22. https://doi.org/10.18524/1810-4215.2020.33.216304

\bibitem{2016TugayDylda} Tugay, A.~V., Dylda, S.~S., \& Panko, E.~A.\ 2016, Odessa Astronomical Publications, 29, 34. https://doi.org/10.18524/1810-4215.2016.29.84961
 
\bibitem{2017ShevchTugay}  Shevchenko, S.~Y. \& Tugay, A.~V.\ 2017, Odessa Astronomical Publications, 30, 45. https://doi.org/10.18524/1810-4215.2017.30.114263

\bibitem{2020Vasylenko} Vasylenko, A.~A., Vavilova, I.~B., \& Pulatova, N.~G.\ 2020, Astronomische Nachrichten, 341, https://doi.org/801.10.1002/asna.202013783

\bibitem{2018Tugay} Tugay, A.~V., Pulatova, N.~G., \& Zhoga, A.~D.\ 2018, Odessa Astronomical Publications, 31, 42. https://doi.org/10.18524/1810-4215.2018.31.144738

\bibitem{2015Pulatova} Pulatova, N.~G., Vavilova, I.~B., Sawangwit, U., et al.\ 2015, MNRAS, 447, 2209. https://doi.org/10.1093/mnras/stu2556

\bibitem{Wang2006A} Wang, T.-G., Zhou, H.-Y., Wang, J.-X., et al.\ 2006, The Astrophysical Journal, 645, 856. https://doi.org/10.1086/504397

\bibitem{Schwartz2006} Schwartz, D.~A., Marshall, H.~L., Lovell, J.~E.~J., et al.\ 2006, The Astrophysical Journal, 640, 592. https://doi.org/10.1086/500102

\bibitem{Struder2001} Str{\"u}der L., Briel U., Dennerl K., Hartmann R., Kendziorra E., Meidinger N., Pfeffermann E., et al., 2001, A\&A, 365, L18. https://doi.org/10.1051/0004-6361:20000066

\bibitem{2018Melnyk} Melnyk O., Elyiv A., Smol{\v{c}}i{\'c} V., Plionis M., Koulouridis E., Fotopoulou S., Chiappetti L., et al., 2018, A\&A, 620, A6. https://doi.org/10.1051/0004-6361/201730479

\bibitem{2015Finet}Finet F., Elyiv A., Melnyk O., Wertz O., Horellou C., Surdej J., 2015, MNRAS, 452, 1480. https://doi.org/10.1093/mnras/stv1401

\bibitem{2012Elyiv}Elyiv A.~A., Hnatyk B.~I., 2007, KPCB, 23, 56. https://doi.org/10.3103/S088459130702002X

\bibitem{2012Pierre}Pierre M., Pacaud F., Adami C., Alis S., Altieri B., Baran N., Benoist C., et al., 2016, A\&A, 592, A1. https://doi.org/10.1051/0004-6361/201526766

\bibitem{2018Koulouridis}Koulouridis E., Ricci M., Giles P., Adami C., Ramos-Ceja M., Pierre M., Plionis M., et al., 2018, A\&A, 620, A20. https://doi.org/10.1051/0004-6361/201832974

\bibitem{2016Vavilova}Vavilova I.~B., Vasylenko A.~A., Babyk I.~V., Pulatova N.~G., 2016, agnw.conf, 105. https://doi.org/10.5281/zenodo.60643

\bibitem{2015Vavilova}Vavilova I.~B., Ivashchenko G.~Y., Babyk I.~V., Sergijenko O., Dobrycheva D.~V., Torbaniuk O.~O., Vasylenko A.~A., et al., 2015, KosNT, 21, 94. https://doi.org/10.15407/knit2015.05.094

\bibitem{2015aVavilova}Vavilova I.~B., Vasylenko A.~A., Babyk I.~V., Pulatova N.~G., 2015, OAP, 28, 150

\bibitem{2016Menzel}Menzel M.-L., Merloni A., Georgakakis A., Salvato M., Aubourg E., Brandt W.~N., Brusa M., et al., 2016, MNRAS, 457, 110. https://doi.org/10.1093/mnras/stv2749

\end{thebibliography}
\end{document}